\documentclass[pdflatex,sn-mathphys-num]{sn-jnl}

\usepackage{graphicx}%
\usepackage{multirow}%
\usepackage{amsmath,amssymb,amsfonts}%
\usepackage{amsthm}%
\usepackage{mathrsfs}%
\usepackage[title]{appendix}%
\usepackage{xcolor}%
\usepackage{textcomp}%
\usepackage{manyfoot}%
\usepackage{booktabs}%
\usepackage{algorithm}%
\usepackage{algorithmicx}%
\usepackage{algpseudocode}%
\usepackage{listings}%

\usepackage{color,natbib,tikz,hyperref}
\newcommand\bm[1]{\mbox{\boldmath $#1$}}
\newcommand\commentout[1]{}

\font\SYM=msbm10 
\def\Real{\hbox{\SYM R}}

\def\npk{\bm{k}}
\def\npl{\bm{l}}

\def\df{{\rm d}}

\newcommand\BL{Boyer and Lindquist}
\newcommand\GM{Garc\'{\i}a-Compe\'an and Manko}
\newcommand\GP{Griffiths and Podolsky}

\begin{document}

\title[Two discs and a missing triangle]{Two discs and a missing triangle: the maximally extended Kerr black hole revisited}

\author*[1]{\fnm{M.A.H.} \sur{MacCallum}}\email{M.A.H.MacCallum@qmul.ac.uk}

\affil*[1]{\orgdiv{School of Mathematical Sciences}, \orgname{Queen Mary University of London}, \orgaddress{\street{Mile End Road}, \city{London}, \postcode{E1 4NS}, \country{UK}}}

\abstract{
The region spanning the ring singularity in the Kerr black hole
solution is described here as two (flat) discs rather than one. From
this viewpoint it follows that the usual depiction of the worldsheet
$\theta=\pi/2$, $\phi$ constant (in Boyer--Lindquist coordinates) is
missing (multiple copies of) the region $r<0$, representable as a
triangle in the usual conformal diagram style.
}


\maketitle

\section{Introduction}
Despite containing two commuting Killing vectors, and therefore
depending on only two essential coordinates, the Kerr black hole solution's
geometry and its physical consequences have been
surprisingly tricky to understand. There have been many papers
exploring various aspects, and advanced textbooks discuss the
solution and its properties.

The ideas in this paper were presented at the CarterFest meeting in
honour of Brandon Carter's 80th birthday\footnote{The slides from the
  talk itself are available at
  \url{https://indico.obspm.fr/event/1346/contributions/838/}. Due to
  personal circumstances this write up has been considerably delayed.}. They
have been expanded and re-organized here for clarity and completeness. I
dedicate this paper to Carter.

I cannot survey all the many aspects of the Kerr solution here: for
some reviews of them see \cite{WilVisSco09}.  My focus is instead quite
narrow. I shall deal only with a couple of aspects of the maximal
analytic extension (MAE) of the Kerr vacuum solution. I shall not add
the electromagnetic, NUT, and cosmological constant parameters which
generalize the original solution as published, and lead to the
Kerr--Newman-NUT-(anti)de Sitter (KNN(A)S) form.

Some of the following remarks do however apply, \emph{mutatis
  mutandis}, to generalizations of the vacuum case, particularly those
with added electromagnetic and NUT parameters.
\GP\ \cite{GriPod09}, Chapter 11, gives a useful account of the
Kerr geometry, its analytic extension and its generalization: as
is depicted there, in Figures 11.10 and 11.11, the KNNS and KNNAS
metrics have substantially different MAEs.

In what follows, I shall for brevity mean by ``the Kerr solution'' the
black hole case where the mass and angular momentum constants $m$
(taken to be positive) and $a$ obey $0<a^2<m^2$.  (Kerr's solution
allows extreme ($m^2=a^2$) and hyperextreme ($a^2>m^2$) subcases.)
The metric to be considered is, in the usual Boyer--Lindquist
coordinates,
\begin{equation}\label{BL}
  \df s^2 = -\frac{Q}{R^2} \left[ \df t - a\sin^2 \theta
  \df \phi \right]^2 + \frac{R^2}{Q}\df r^2 + \frac{\sin^2\theta}{R^2}\left[
      a \df t - (r^2 +a^2) \df \phi \right]^2 + R^2 \df \theta^2
\end{equation}
where
\begin{eqnarray}
R^2 &=& r^2 + a^2\cos^2 \theta \\ 
Q &=& r^2 - 2mr + a^2.
\end{eqnarray}
The \BL\ chart neighbourhood on which the coordinates of
Eq.~\eqref{BL} apply can be taken to be in 1-1 correspondence with the
region in $\Real^4$ with coordinates $(t,\,r,\,\theta,\,\phi)$ defined
by $-\infty<t<\infty$, $-\infty<r<\infty$, $0\leq\theta\leq\pi$, and
$0\leq\phi<2\pi$. The bounds on $\theta$ and $\phi$ are chosen so that
if $a=0$ (the Schwarzschild solution), then as $r\rightarrow \infty$,
$\theta$ and $\phi$ become the standard spherical polar coordinates on
a sphere at infinity.  As usual for such coordinates, $\phi=2\pi$ is
identified with $\phi=0$: this breaks the strict definition of a chart
neighbourhood, but is common practice when there is an ignorable
coordinate on circles.  One may instead of $\theta$ use
$\theta'=\theta-\pi/2$, so that $\theta'\in(-\pi/2,\,\pi/2)$.

The metric \eqref{BL} on this chart neighbourhood is clearly regular
everywhere except at $r=0$, $\theta=\pi/2$, where $R=0$, and at the
horizons where $Q=0$, across which, as it turns out, the spacetime can
be analytically extended (i.e.\ $Q=0$ gives coordinate singularities,
not curvature singularities). It has a zero Ricci tensor and a Weyl
curvature tensor of Petrov type D such that in a Newman--Penrose frame
\citep{NewPen62} for which $\npk$ and $\npl$ are chosen to be the two
repeated principal null directions (PNDs), only
$\Psi_2=-m/(r+ia\cos\theta)^3$ is non-zero.

By the Kundt--Thompson theorem, the congruences defined by the PNDs are
geodesic and shearfree. The Weyl curvature is unbounded as
$r\rightarrow 0$ and $\theta\rightarrow\pi/2$ (note that $\Psi_2$ is a
Cartan invariant \citep{BroChaCol18} which locates the singularity
without the ambiguities in locating it via the Kretschmann scalar
\cite{ChrMalYun20}). $R=0$ is therefore a curvature singularity: its
points are nonetheless ordinary points of the chart neighbourhood. It
is described as a ring singularity because in various coordinate
systems its points are mapped to a ring in $\Real^4$: for example in
Weyl's canonical coordinates $(t,\,\rho,\,z,\,\phi)$ for the Kerr
solution, where $t$ and $\phi$ are as in \BL\ coordinates but
\begin{equation}\label{BL2Weyl}
\rho=\sqrt{r^2-2mr+a^2}\sin \theta, \qquad z=(r-m)\cos\theta,
  \end{equation}
  the singularity is at $\rho=a$ (and $z=0$). Weyl coordinates form one of the
  several coordinate systems in which the Kerr solution has been
  written (see Section 20.5 of \cite{SteKraMac03}, and
  \cite{Car66,Dor00,Vis09,BaiBerSim21} for other examples).

Calling this a ring could be thought mildly misleading: as
Visser \cite{Vis09} says, the singularity's intrinsic geometry is of course
singular.  A much fuller investigation of the singularity was
made in ref.\ \cite{ChrMalYun20}, see also \cite{Pun90}.

The additional constant parameters in the more general KNN(A)S cases
appear, together with $m$ and $a$, in the Cartan invariants
characterizing the geometries: see \citep{BroChaCol18} which
clarified and extended earlier work of Abdelqader and Lake
\cite{AbdLak13} and Page and Shoom \cite{PagSho15}. Setting those
parameters to zero gives the invariants for the Kerr
solution, for which  Brooks {\em et al.} \citep{BroChaCol18} showed that the black
hole event horizon, usually denoted $r_+:=m+\sqrt{m^2-a^2}$, the inner
horizon $r_-:=m-\sqrt{m^2-a^2}$, the stationary limit surface
i.e.\ the outer boundary of the ergosphere\footnote{As Visser \cite{Vis09}
remarks, one has to beware that authors use the terms ergosphere and
ergoregion in varied ways.},
$r_{e^+}:=m+\sqrt{m^2-a^2\cos^2\theta}\geq r_+$, and the inner
stationary limit surface $r_{e^-}:=m-\sqrt{m^2-a^2\cos^2\theta}\leq
r_-$ (which meets the singular points where $R=0$) all appear as zeros
of invariants. At the horizons the Killing vector $\partial_t$, which
is timelike in $r>r_+$ where the metric is stationary, becomes null,
and between them $t$ is a spacelike coordinate. Indeed $t$ is
spacelike in the larger ergoregion $r_{e^-}<r<r_{e^+}$, except at the
horizons.

Carter was the first to give the complete maximal analytic extension
of the symmetry axis of the Kerr solution \citep{Car66}. A short time
later he gave a full discussion of the global structure of the Kerr
family \citep{Car68}, following on from and agreeing with that of Boyer and
Lindquist \cite{BoyLin67}. See also \cite{Car73} and \cite{ONe95}.  Inevitably,
any new material below is only a small addendum to the discussions of
Carter and of \BL\ (and later work).

Reading the papers of \GM\ \cite{ManGar14,GarMan15}, which critique
and extend earlier discussions, prompted me to revisit the MAE of the
Kerr solution, stimulating my remarks below.

\section{Carter's paper of 1966}

In his 1966 paper, Carter considered the two-dimensional space of
the axis of the Kerr solution (i.e.\ $\theta=0$ in
Eq.~\eqref{BL}). The metric in this space can be written
\begin{equation}\label{axis}
  \df s^2 = -\left(1-\frac{2mr}{r^2+a^2}\right) \df u^2 + 2\df u\, \df r
\end{equation}
with a null coordinate $u$. This metric is analytic and nonsingular for
all $(u,\,r)$. 

\begin{figure}[ht]
\begin{center}
  \includegraphics[scale=1,angle=90]{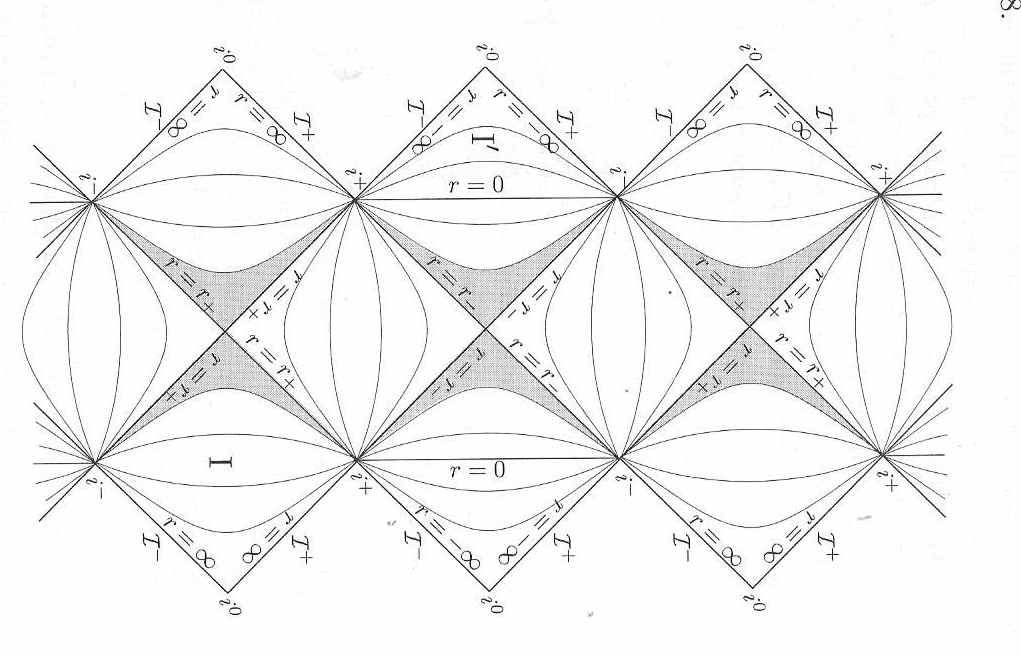}
  \end{center}
\caption{The conformal diagram of the maximally extended Kerr black
  hole for a worldsheet on which $\theta(\neq \pi/2)$ and
  $\phi$ are constant. The shaded areas indicate
  the ergoregions. Figure taken from \cite{GriPod09}, Fig.\ 11.8, which
  is itself after \cite{Car66}.}
  \label{Fig1}
\end{figure}

Considering the null geodesics in this space, Carter found that it is
geodesically incomplete as $r\rightarrow r_+$ and $u\rightarrow
-\infty$ and as $r\rightarrow r_-$ and $u\rightarrow \infty$,
i.e.\ where $Q=0$. A
further coordinate transformation introducing a second null coordinate
$w$ was used to show that an analytic continuation could be made to 3 regions
joined at the lines $r=r_\pm$. Carter then noted that by repeating
the arguments one could
analytically continue the solution into multiple copies of the
original region, joined across $r=r_\pm$ as in the by-now familiar
diagram, Fig.\ \ref{Fig1}. The diagram applies equally well to the two-surface
for constant $\phi$ and any $\theta \neq
\pi/2$. 

In this figure, the intersection of the axis worldsheet and a
particular \BL\ chart neighbourhood consists of the region ${\rm I}'$ and the
diamond-shaped regions above and to its right, including the part
where $r <0$.  Its boundaries are at infinity, i.e.\ $|r|\rightarrow
\infty$, and, less obviously, at the inner and outer horizons, where
analytic continuation is possible.  The metric in $r<0$ is unaltered
by the coordinate map $r\rightarrow r'=-r$ coupled with a change of
sign of $m$. The whole \BL\ chart neighbourhood with metric \eqref{BL}
can thus equally well be regarded as a Kerr solution for $m<0$.

Carter's diagram (Fig.\ \ref{Fig1} above), in which conformal transformations
are used to represent infinite regions by finite regions in
the plane, is like the conformal diagrams introduced by
Penrose \cite{Pen64}: such diagrams are often now called Penrose-Carter (or
Carter-Penrose) diagrams. The portions of the figure shown as squares
with sides at $45^o$ to the lines of text have been called Carter
blocks by Zannias \cite{Zan18}.  Since the figure contains multiple isometric
regions ``there are several ways in which different parts may be
identified'' \citep{Car66}.

\section{Surfaces of constant \BL\ coordinates}
\label{BLsec}
Kerr and Schild \citep{KerSch65,KerSch65a} had considered metrics of
the form $g_{ab}=\eta_{ab}+2Hk_ak_b$ where $\eta_{ab}$ is the metric
of flat Minkowski space and $k_a$ is a vector null in both $g_{ab}$
and $\eta_{ab}$ (the resulting solutions of the Einstein equations are
discussed in Chapter 32 of \cite{SteKraMac03}).  Boyer and Lindquist
quote the Kerr-Schild form of the Kerr metric, which was Kerr's
original form, using coordinates $\{x,y,z,t\}$ which are the usual
pseudo-Cartesian coordinates in the Minkowski part of the metric. They
call this the $M$-system (their Eq.\ (2.4)). For this form $k_a$ is
one of the PNDs of the Kerr solution. \BL\ denote the similar
coordinate system based on the other PND by $M'$.

\BL\ then pass to a form they describe as ``a generalization of the
Eddington form of the Schwarzschild metric'', which they call the (E) frame
(their (2.7)). Later they transform again to what they call
``Schwarzschild-like coordinates'' (their (2.13)), which are now known
as Boyer-Lindquist coordinates and in which the metric is \eqref{BL}.

Boyer and Lindquist discuss in their Sect.\ 2 a pictorial
representation of surfaces of constant $r$, surfaces of constant
$\theta$ (or $\theta'$), and surfaces of constant $\phi$, all at constant $t$, using
coordinates $(x,\,y,\,z)$ obtained from the coordinate transformation to
the M frame, and represented as Cartesian. Note that these surfaces
are diffeomorphic to but in general not isometric to the corresponding
surfaces in the Kerr geometry. In this representation, the constant $r$
surfaces are confocal ellipsoids. (For $3m^2/4<a^2<m^2$ this
is misleading in that the Gaussian curvature of the event horizon
becomes negative near the poles $\theta'=\pm \pi/2$
\citep{Sma73,Vis09}.)

The constant $\theta$ surfaces (other than the axis $\theta=0$) are
represented by half-hyperboloids of one sheet, confocal to the
ellipsoids. The ``half'' arises because \BL, at this point in their
paper, are tacitly assuming that $r>0$, so the coordinate
transformation formula $z=r\cos \theta$ implies that $z$ and $\cos
\theta$ have the same sign.  The constant $\theta$
surfaces lie in $z\gtrless 0$ according as $\theta\lessgtr \pi/2$.

The constant $\phi$ surfaces ``have the appearance of bent
planes, which are approximately vertical for large $r$ but flatten out
and become horizontal at the edge of the disc'' and \BL\ show these are
ruled quartic surfaces obtainable from one another by rotation about
the $z$ axis.

Apropos of the limit $r\rightarrow 0$, \BL\ say that ``the ellipsoid
degenerates to a disk'', $x^2+y^2 = a^2 \sin^2 \theta$, $z = 0$, and
that ``the boundary of this disk [is the] set of points at which the
metric (2.7) becomes singular.'' Note that their argument only shows
that this two-surface is homeomorphic to the region of the Euclidean
plane for which, in Cartesian coordinates, $x^2+y^2 \leq a^2$ for some
fixed $a$, i.e.\ that it is topologically a disc. It does not show that the
two-surface is isometric to a Euclidean disc. To avoid confusion of
the two ideas I shall use the terms ``topological disc'' and ``flat
disc'' respectively.  

Using the specialization of Eq.~\eqref{BL} to $r=0$, $t=$constant, the
metric on the disc $r=0$ is
\begin{equation}\label{flatdisc}
  \df s^2 = a^2 \sin^2 \theta\, \df \phi^2 + a^2 \cos^2 \theta\, \df \theta^2
\end{equation}
which, defining $\chi=a\sin \theta$, can be written as $ \df \chi^2 +
\chi^2\df \phi^2$, the usual metric of a flat disc in polar
coordinates. 

\section{The disc(s) spanning the ring singularity}
\label{discs}

A detail \BL\ do not discuss is that taking the range
$\theta\in(0,\pi)$ (excluding $\theta=\pi/2$)
covers all values of $\chi\in[0,a)$ twice,
once with $0<\theta<\pi/2$ and once with $\pi/2<\theta<\pi$. Thus
$r=0$ consists of two isometric discs, each spanning the ring
singularity and each continuable to the region $r<0$, rather than
one. Each has the singular ring as its boundary.

The two discs appear in a number of the discussions of the Kerr
solution, but are not always clearly so described. The earliest I
am aware of\footnote{I thank Vladimir Manko drawing this to my
  attention,} is for the hyperextreme case, rather than that under
discussion here, and appears as Fig.~27 of \cite{HawEll73}: the
diagram there is essentially identical to Fig.\ \ref{GarMan1} below,
with, of course, a different caption.

\begin{figure}
  \begin{center}
  \includegraphics[scale=0.3]{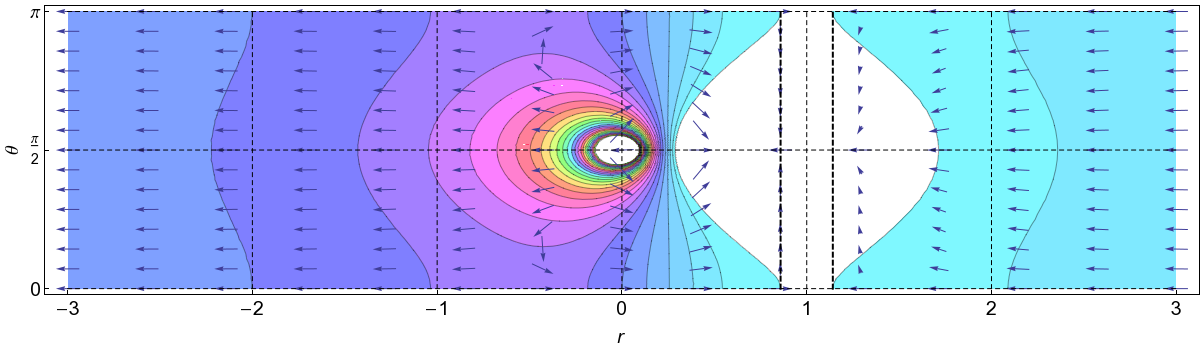}
  \end{center}
  \caption{A representation of the $(r,\theta)$ plane in
    \BL\ coordinates (for $|r|\leq 3$) for a solution with $m=1$. This
    figure (originally plotted for a Kerr--Newman solution but the Kerr
    black hole will give a similar diagram) is taken from an anonymous
    referee report, with the originator's permission. The white area
    on the right indicates the stationary limits and the region
    between the horizons. Note that the equivalent diagram for $m=-1$
    is the same as this but reversed left-right.}
  \label{RTheta}
\end{figure}

  The two discs' occurrence is
particularly clear in Fig.\ \ref{RTheta} (this is for a Kerr--Newman
solution but the Kerr solution figure is analogous).
In this figure the ring singularity is represented by the point $r=0$,
$\theta=\pi/2$ and the two discs are represented by the rest of the
line $r=0$.

Although the presence of two discs has been stated for the vacuum Kerr
context, it applies to the KNN(A)S generalizations, the differences
lying in the curvature of the (topological) disc(s) \citep{GarMan15}.
For example, the two discs can be seen, for the Kerr--Newman
solution, in Fig.~3 of \cite{GarMan15}, where $r=0$, for particular
$m$, $a$ and charge $Q$, is depicted in cylindrical polar
coordinates: one disc appears as the surface of revolution generated by
the arc in $z\in(-2,0)$ and the other as that of the arc in
$z\in(0,2)$.  The corresponding arcs in the $(\rho,\,z)$ plane for the
Kerr solution, i.e.\ in Weyl coordinates, can be parametrized as $\rho = a\sin\theta$,
$z=-m\cos\theta$.

\GM\ \cite{GarMan15}, having shown that $r=0$ in the Kerr solution is flat,
argue that the only flat two-surface it can be is a dicone, i.e.\ a
pair of isometric cones.  Their conclusion is obtained after taking
the negative mass interpretation and using the Weyl coordinates of
Eq.~\eqref{BL2Weyl}. \GM\ say ``it is obvious that the disk geometry must
be excluded because the coordinate $z$ in (7) runs over all values in
the interval $[-|M|, +|M|]$, while $z$ in the case of a disk must take
only one particular value''. This argument is correct in flat space
but not, as shown below, in the Kerr metric.

The distinction between a flat disc and a cone with the same central
point (the centre of the disc or the apex of the cone) lies in the
ratio of the length $\ell$ of a circle of revolution and its
distance $d$ from the central point. For a flat disc $\ell/d$ will be
$2\pi$; for a cone it will be smaller.

In flat space, consider a surface of revolution given by rotating a
curve $z=f(\rho)$ about the $z$-axis of Weyl
coordinates, and let $z_0=f(0)$. If $f$ is constant the distance of a
point on this curve from the axis is simply $\rho$ and the region
$z=z_0$, $\rho^2\leq a^2$ is a flat disc. If $f$ is not constant, a
circle radius $\rho$ about the $z$-axis will be at a distance in the
surface of $\int_0^\rho \sqrt{1+f'^2}\,\df \rho$ from $\rho=0$, $z=z_0$,
where $f'=\partial f/\partial \rho$. The ratio of the circle's length
to this distance will be less than $2\pi$ unless $f'=0$. So \GM's
remark is correct in flat space.

In the metric \eqref{BL}, the topological disc $r=0$ at constant $t$ with
$0\leq \theta<\pi/2$ has its centre at $\theta=0$ and the length of the
circle at $\theta=\theta_0$ is $2\pi a \sin
\theta_0$, while the radial distance of this circle from $\theta =0$ is $\int_0^{\theta_0}
a \cos \theta \df \theta= [a \sin \theta]_0^{\theta_0}=a \sin
\theta_0$. The disc is therefore a flat disc, not a cone. The same
clearly applies also to  the disc $\pi/2 < \theta \leq \pi$.

One can readily, using \eqref{BL2Weyl}, reach the same result using Weyl coordinates.

\commentout{
The general Weyl coordinate form for a metric is
\begin{equation}
\df s^2  =  {\rm e}^{-2U}[{\rm e}^{2k}(\df \rho ^2+\df z^2)+
\rho ^2\df \varphi ^2]-{\rm e}^{2U}\df t^2.
\end{equation}
From  Eq.~\eqref{BL2Weyl} one finds that
\begin{equation}
\df \rho^2 +\df z^2 = (r^2-2mr+m^2\sin^2\theta+a^2\cos\theta)(\df
r^2/Q+\df \theta^2, \end{equation} so comparison with the
two-dimensional metric induced on a surface of constant $t$ and $\phi$
by \eqref{BL} shows that $$ e^{-2U}e^{2k} =
(r^2+a^2\cos^2\theta)/(r^2-2mr+m^2\sin^2\theta+a^2\cos\theta).$$ In a
(topological) disc $r=0$ this is
$a^2\cos^2\theta/(m^2\sin^2\theta+a^2\cos\theta)$ and from
Eq.~\eqref{BL2Weyl} a radial arc in this disc has $\rho =
a\sin\theta$, $z=-m\cos\theta>0$

 A point  $\rho_0 = a\sin\theta_0$, $z_0=-m\cos\theta_0$ on that arc
 is at a distance in the surface
of $d=\int_{\pi/2}^{\theta_0} \sqrt{a^2\cos^2\theta(\df \rho^2+\df
z^2)/(m^2\sin^2\theta+a^2\cos\theta)}$ from $\rho=0$, $z=|m|$, and
using the parametrization of the curve by $\theta$ this is easily seen
to be $\rho_0$. Hence $r=0$ in the Kerr solution has two flat discs
spanning the ring singularity, not a dicone.}

It is of course trivial to make a flat disc into a cone, just by
changing the period of the $\phi$ coordinate to $p<2\pi$, but that
would mean there was non-zero holonomy round circles $\phi\in(0,p)$
throughout the spacetime.  (Indeed one can also make $p>2\pi$, or even infinite.)

In the Kerr--Newman solution (see \cite{GarMan15}) and the more general
solutions discussed in \cite{ManGar14}, the two-surface(s) $r=0$
spanning the ring singularity are not flat discs, although they are
still topological discs (\GM\ say they are not discs, but clearly mean
flat discs).  They are curved disc portions of a sphere or
pseudosphere, depending on the curvature.

Manko and Garc{\'i{\i}}a-Comp\'ere \cite{ManGar14} say ``Taking into account that the $r=0$, $t =$ const
surface of the Kerr solution ($\Lambda = 0$) is a dicone..., it would
be plausible to suppose that in the Kerr–anti–de Sitter case the
respective surface is represented by some conic surface of a constant
negative Gaussian curvature with singular vertices and
equator....''. Since the premise here is mistaken, it instead seems plausible
that the curved (topological) discs in these cases do not have
cusps. This could be verified by an argument analogous to the one
above.

\section{A \BL\ chart neighbourhood}

The exterior of the Kerr black hole is given by Eq.~\eqref{BL} with $m>0$
in the region $r>r_+$.  Many discussions focus on the analytic
extension of the black hole to and through $r=0$ (rather than the
continuations across horizons). These discussions provide ways of
describing the whole of a \BL\ chart neighbourhood for the Kerr black
hole. As \BL\ correctly say ``continuation [at the disc] to negative $r$
values is permissible because the metric [i.e.\ Eq.~\eqref{BL}] remains
regular at $r = 0$''. One could simply identify the two discs with one
another, but, as \GP\ \cite{GriPod09} remark, such an extension is not
C${}^1$ (as is shown in \cite{ChrMalYun20}).

One way to describe the extension to $r<0$  is given by Fig.\ 1 of \cite{GarMan15},
reproduced here.
\begin{figure}[ht]
  \begin{center}
    \includegraphics[scale=1]{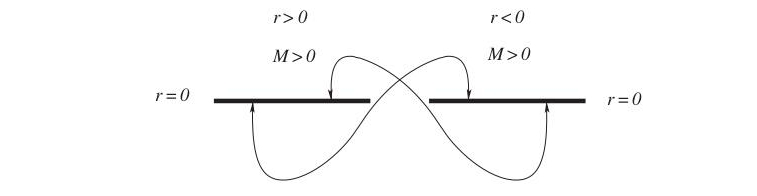}
  \end{center}
  \caption{Figure 1 of \cite{GarMan15}.}
  \label{GarMan1}
\end{figure}
This figure does depict the two discs described above. As
\GM\ \cite{GarMan15} note, the ``upper-bottom'' identification shown is
required so that the curve depicted joins $r>0$, $\theta=\psi$ with
$r>0$, $\theta=\pi-\psi$. That curve in Fig.\ \ref{GarMan1} could also be
drawn as a closed curve in Fig.\ \ref{RTheta} enclosing the
singularity at $r=0$, $\theta=\pi/2$. This is discussed further in Sect.\ \ref{triangle}.

Figure \ref{GarMan1} is essentially the same as the one in  Hawking and Ellis \cite{HawEll73} mentioned above. However, the hyperextreme solution
which was considered by Hawking and Ellis has no horizons and the MAE
is covered by a single \BL\ chart. Thus Fig.\ \ref{GarMan1} can represent
the complete MAE for that case.  The caption to the similar figure
appearing as Fig.\ 11.2 in \GP\ \cite{GriPod09} (who cite the discussion in
\cite{HawEll73}), also says the diagram shows a MAE for the Kerr
solution, but in the black hole case the full MAE also requires the
continuations across the horizons found by Carter.

The fact that the region $r<0$ can be interpreted as the $r>0$ region
of a Kerr solution for $m<0$ has led to descriptions in which two Kerr
solutions with $r>0$, one for $m>0$ and one for $m<0$, are glued
together across the discs at $r=0$ \citep{GriPod09,GarMan15}. Such a
description amounts to cutting the \BL\ chart neighbourhood along the
two discs, using the changed coordinate $r'=-r$ in the $r<0$ region,
and then rejoining the two parts.  While perfectly correct as a
description of an entire \BL\ chart neighbourhood, this description
is misleading insofar as it obscures the fact that the Boyer-Lindquist
coordinate neighbourhoods for the $m\gtrless 0$ solutions are one and
the same, just with the $r$ coordinate replaced by $r'$. (Strictly,
the presentation of the $m<0$, $r'>0$, region is using a different
chart on the same chart neighbourhood, i.e.\ a different map of the
chart neighbourhood into $\Real^4$.)

The idea of gluing metrics \eqref{BL} with positive and negative $m$
is accompanied in \cite{GarMan15} by a puzzling picture (\GM's
Fig.\ 2) which depicts an $m<0$ ring singularity as outside the event
horizon and stationary limit, and an $m>0$ ring singularity as inside,
whereas these rings are one and the same. There are no horizons in the
$r<0$ region, and, after the coordinate change to $r'$, the horizons
are at negative values of $r'$, $r'_+<r'_-<0$, and so at smaller $r'$
than that of the ring, $r'=0$. (The puzzling figure depicts the
horizons at a positive radius.)

\section{The maximal analytic extension and its missing triangle}\label{triangle}

To present the full MAE in a good way is not simple. It helps that the angular
coordinate $\phi$ is ignorable, and all diagrams in the literature do
ignore it, so each point shown is really a circle of points. I shall
follow the same practice here.  Following  Carter \cite{Car66}, the temporal and
radial structure is usually presented in two-dimensional conformal
diagrams of worldsheets with one space and one time dimension, such as
Fig.\ \ref{Fig1}: these are surfaces of constant $\theta$ and $\phi$.

In Carter's later paper \cite{Car68}, which appeared after that of \BL, the global
structure for the Kerr--Newman family is considered. By adaptation and
generalization of the arguments in \cite{Car66}, Carter gave the
more complicated coordinate transformations required to find the
four-dimensional maximal analytic extension (MAE).

That paper also integrates the geodesics using the Carter constant, a
quadratic found by considering separability of the Hamilton--Jacobi
equations (see also \cite{Car68,Car73}), which was later shown to
arise from a Killing tensor \citep{WalPen70}.

Carter could then show that all geodesics in the extended manifolds
that do not reach the singularities (plural here due to the infinitely
many repeats of the basic blocks) are complete, and that causal
geodesics that do reach the singularities lie in the equatorial planes
\cite{Car68} (cf.\ \cite{Pun90}). The paper did not include diagrams
like those of the 1966 paper, which had by then been generalized by
\BL.

Carter further showed that in a region near the singular ring the
periodic Killing vector $\partial_\phi$ is timelike, so that there the
circles $0\leq \phi \leq 2\pi$ are closed timelike lines and imply
causality violation. For more on that and the behaviour of geodesics
see \cite{ChrMalYun20}.

\begin{figure}[ht]
  \begin{center}
    \includegraphics[angle=90, width=0.8\textwidth]{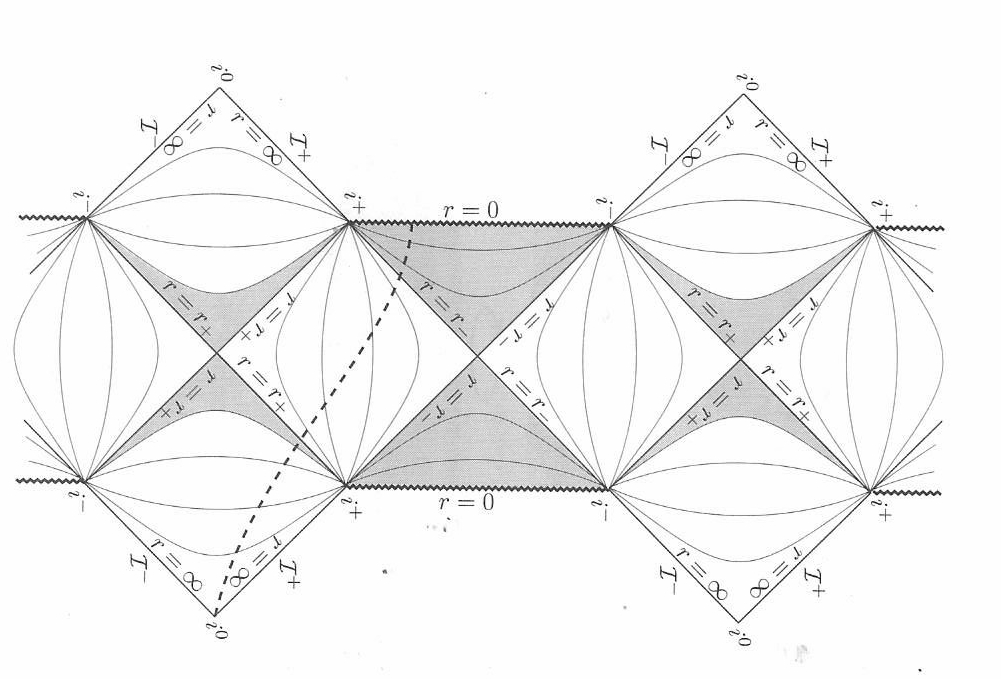}
  \end{center}
  \caption{The maximal analytic extension of $\theta=\pi/2$, $\phi$
constant, as shown in Figure 11.7 of
\cite{GriPod09}.}
  \label{GP117}
\end{figure}

The well-known conformal diagram for the maximal analytic
extension of $\theta=\pi/2$, $\phi$ constant,  is shown here as Fig.\ \ref{GP117}.

However, this picture does not give all the points with $\theta=\pi/2$
in the MAE of the Kerr solution.  To see that there are missing
points, reconsider a closed curve in Fig.\ \ref{RTheta} encircling the
singularity at $r=0$, $\theta=\pi/2$. One can start at $r>r_+$,
$\theta=\theta_0>\pi/2$ and let $r$ decrease along the curve, keeping
$\theta$ constant, cross the disc $r=0$, $\theta \in (\pi/2,\,\pi)$,
then at (say) $r=-2$ let $\theta$ reduce to a value in $(0, \pi/2)$,
and finally let $r$ increase again at constant $\theta$ to $r>r_+$,
followed by increasing $\theta$ to $\theta_0$.  Another depiction of
such a curve is as the path shown in Fig.\ \ref{GarMan1}.  Note that
although there are paths in a \BL\ chart between any point with $r>0$
and $\theta=\pi/2$ and any point with $r<0$ and $\theta=\pi/2$, they
have to pass through $r=0$ at some value of $\theta \neq \pi/2$: there is no
such path with $\theta=\pi/2$ at all points, since $\theta=\pi/2$,
$r=0$ is singular.

\begin{figure}[ht]
  \begin{center}
    \includegraphics[scale=0.8]{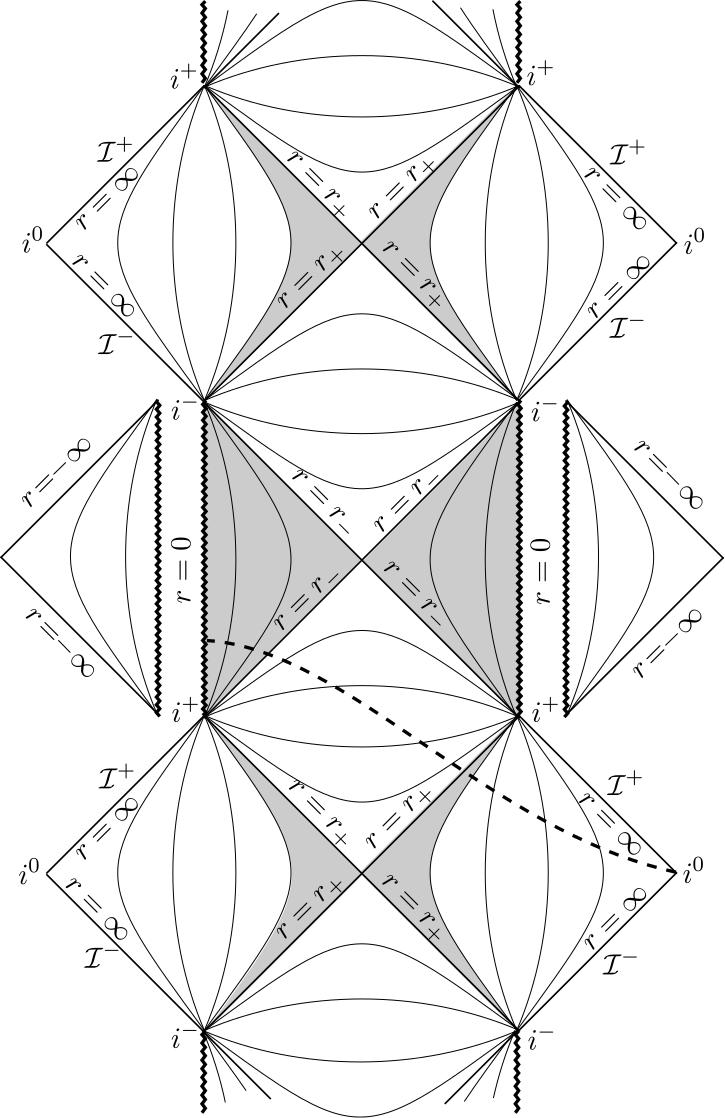}
  \end{center}
  \caption{Fig.\ \ref{GP117} modified by the addition of two of the missing
    triangles.}
  \label{Fig5}
\end{figure}

The missing points are those with $\theta=\pi/2$, $r<0$, shown as a
line in Fig.\ \ref{RTheta}. They can be depicted in exactly the same way as the
triangle in region ${\rm I}'$ of Fig.\ \ref{Fig1} for which $r<0$, except that
for $\theta=\pi/2$ the boundary at $r=0$ of this region is
singular. This is the ``missing triangle'' of my
title. Figure \ref{Fig5} shows the usual conformal diagram for
$\theta=\pi/2$, $\phi$ constant, corrected by the addition of two of
the copies
of this triangle.

To obtain a three-dimensional representation of the MAE of the Kerr
black hole (still suppressing $\phi$) one can stack copies of
Fig.\ \ref{Fig1}, one for each value of $\theta\neq \pi/2$, together
with one copy of Fig.\ \ref{Fig5} for $\theta=\pi/2$. These stacked
figures intersect a particular one of the multiply-repeated \BL\ chart
neighbourhoods along lines $\theta=$ constant in an $(r,\,\theta)$
plane as depicted in Fig.\ \ref{RTheta}.

\section{Further remarks}
It has been clarified above that the two-surfaces spanning the ring
singularity in the Kerr black hole solution are a pair of discs, not
just one, and both are flat, and it has also been shown that the usual
conformal diagram for the $\theta=\pi/2$ section of the MAE is missing
(multiple copies of) a triangle of points with $r<0$ (in
\BL\ coordinates).

A number of other recent developments relate to the issues discussed
above, and deserve mention to interested readers. Chru\'sciel et
al. \cite{ChrOlzSzy12} introduced ``projective diagrams'' as an
alternative way to depict the analytic extensions and the geometry of
the Kerr family: these were presented in Chru\'sciel's talk at the
Carter Fest. Podolsk\'y and Vr\'atn\'y \cite{PodVra21} have presented
a new metric form for all Petrov type D solutions. In his talk at the
CarterFest, Riazuelo showed some enjoyable computer graphics of the
Kerr family MAEs (cf.\ \cite{Ria20}.)

It is remarkable that 54 years on from Carter's seminal 1966 paper, the MAE
of the Kerr solution remains an object of study prompting new work.

\backmatter

\bmhead{Acknowledgements}

I am grateful to Piotr Chru\'sciel, Tomas Ledvinka and Jiri Podolsky
for inputs which helped me improve this paper compared with the original
talk as given at the Carter Fest, to Griffiths and Poldolsky for help
with the figures from their book, and to Podolsky for encouragement to
complete and submit this version.

\section*{Declarations}

\bmhead{Conflict of interest}
The author declares no conflict of interest.

\bibliography{CarterFest}

\end{document}